# Performance Evaluation of Wimax Physical Layer under Adaptive Modulation Techniques and Communication Channels

Md. Ashraful Islam
Dept. of Information & Communication Engineering
University of Rajshahi, Rajshahi, Bangladesh
e-mail: ras5615@gmail.com

*Riaz Uddin Mondal* (corresponding author)
Assistant Professor, Dept. of Information & Communication Engineering
University of Rajshahi, Rajshahi, Bangladesh
e-mail: riaz_uddin_ru@yahoo.com

*Md. Zahid Hasan*
Dept. of Information & Communication Engineering
University of Rajshahi, Rajshahi, Bangladesh
e-mail: hasan.ice@gmail.com

*Abstract*— **Wimax (Worldwide Interoperability for Microwave Access) is a promising technology which can offer high speed voice, video and data service up to the customer end. The aim of this paper is the performance evaluation of an Wimax system under different combinations of digital modulation (BPSK, QPSK, 4-QAM and 16-QAM) and different communication channels AWGN and fading channels (Rayleigh and Rician). And the Wimax system incorporates Reed-Solomon (RS) encoder with Convolutional encoder with ½ and 2⁄3 rated codes in FEC channel coding. The simulation results of estimated Bit Error Rate (BER) displays that the implementation of interleaved RS code (255,239,8) with 2/3 rated Convolutional code under BPSK modulation technique is highly effective to combat in the Wimax communication system. To complete this performance analysis in Wimax based systems, a segment of audio signal is used for analysis. The transmitted audio message is found to have retrieved effectively under noisy situation.**

I. INTRODUCTION

The demand for broadband mobile services continues to grow. Conventional high-speed broadband solutions are based on wired-access technologies such as digital subscriber line (DSL). This type of solution is difficult to deploy in remote rural areas, and furthermore it lacks support for terminal mobility. Mobile Broadband Wireless Access (BWA) offers a flexible and cost-effective solution to these problems [1].

The IEEE WiMax/802.16 is a promising technology for broadband wireless metropolitan area networks (WMANs) as it can provide high throughput over long distances and can support different qualities of services. WiMax/802.16 technology ensures broadband access for the last mile. It provides a wireless backhaul network that enables high speed Internet access to residential, small and medium business customers, as well as Internet access for WiFi hot spots and cellular base stations [2]. It supports both point-to-multipoint (P2MP) and multipoint-to-multipoint (mesh) modes.

WiMAX will substitute other broadband technologies competing in the same segment and will become an excellent solution for the deployment of the well-known last mile infrastructures in places where it is very difficult to get with other technologies, such as cable or DSL, and where the costs of deployment and maintenance of such technologies would not be profitable. In this way, WiMAX will connect rural areas in developing countries as well as underserved metropolitan areas. It can even be used to deliver backhaul for carrier structures, enterprise campus, and Wi-Fi hot-spots. WiMAX offers a good solution for these challenges because it provides a cost-effective, rapidly deployable solution [3].

Additionally, WiMAX will represent a serious competitor to 3G (Third Generation) cellular systems as high speed mobile data applications will be achieved with the 802.16e specification.

The original WiMAX standard only catered for fixed and Nomadic services. It was reviewed to address full mobility applications, hence the mobile WiMAX standard, defined under the IEEE 802.16e specification. Mobile WiMAX supports full mobility, nomadic and fixed systems [4]. It addresses the following needs which may answer the question of closing the digital divide:

- It is cost effective.
- It offers high data rates.
- It supports fixed, nomadic and mobile applications thereby converging the Fixed and mobile networks.
- It is easy to deploy and has flexible network architectures.
- It supports interoperability with other networks.
- It is aimed at being the first truly a global wireless broadband network.

IEEE 802.16 aim to extend the wireless broadband access up to kilometers in order to facilitate both point-to-point and point-to-multipoint connections [5].

## II. SIMULATION MODEL

This structure corresponds to the physical layer of the WiMAX/IEEE 802.16 WirelessMAN-OFDM air interface. In this setup, The input binary data stream obtained from a segment of recorded audio signal is ensured against errors with forward error correction codes (FECs) and interleaved..

The complementary operations are applied in the reverse order at channel decoding in the receiver end. The complete channel encoding setup is shown in above Figure 1.

FEC techniques typically use error-correcting codes (e.g., RS, CC) that can detect with high probability the error location. These channel codes improve the bit error rate performance by adding redundant bits in the transmitted bit stream that are employed by the receiver to correct errors introduced by the channel. Such an approach reduces the signal transmitting power for a given bit error rate at the expense of additional overhead and reduced data throughput (even when there are no errors) [6]. The forward error control (FEC) consists of a Reed-Solomon (RS) outer code and a rate-compatible Convolutional Code (CC) inner code. A block Reed Solomon (255,239,8) code based on the Galois field GF $(2^8)$ with a symbol size of 8 bits is chosen that processes a block of 239 symbols and can correct up to 8 symbol errors calculating 16 redundant correction symbols. Reed Solomon Encoder that encapsulates the data with coding blocks and these coding blocks are helpful in dealing with the burst errors [7]. The block formatted (Reed Solomon encoded) data stream is passed through a convolutional interleaver. Here a code rate can be defined for convolutional codes as well. If there are k bits per second input to the convolutional encoder and the output is n bits per second, the code rate is k/n. The redundancy is on not only the incoming k bits, but also several of the preceding k bits. Preceding k bits used in the encoding process is the constraint length m that is similar to the memory in the system [8], where k is the input bits and n is the number of output bits – is equal to ½ and 2/3 and the constraint length m of 3 and 5. The convolutionally encoded bits are interleaved further prior to convert into each of the either four complex modulation symbols in BPSK, QPSK, 4-QAM, 16-QAM modulation and fed to an OFDM modulator for transmission. The simulated coding, modulation schemes and also noisy fading channels used in the present study is shown in Table 1.

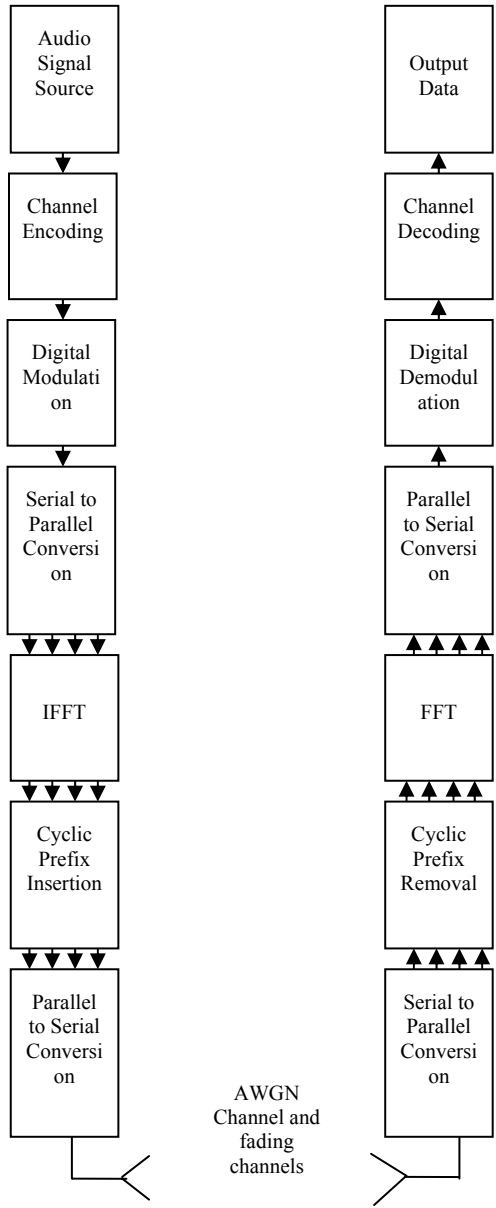

**Figure 1: A block diagram represents Wimax communication system with interleaved concatenated channel coding.**

**Table 1: Simulated Coding, Modulation Schemes and noisy channels**

| Modulation | RS code | CC code rate | Noise Channels |
|---|---|---|---|
| BPSK QPSK 4-QAM 16-QAM | (255,239,8) | (1/2, 2/3) | AWGN Chnnel |
| BPSK QPSK 4-QAM 16-QAM | (255,239,8) | (1/2, 2/3) | Rayleigh Channel |
| BPSK QPSK 4-QAM 16-QAM | (255,239,8) | (1/2, 2/3) | Rician Channel |

In OFDM modulator, the digitally modulated symbols are transmitted in parallel on subcarriers through implementation as an Inverse Fast Fourier Transform (IFFT) on a block of information symbols followed by an analog-to-digital converter (ADC). To mitigate the effects of inter-symbol interference (ISI) caused by channel time spread, each block of IFFT coefficients is typically presented by a cyclic prefix. At the receiving side, a reverse process (including deinterleaving and decoding) is executed to obtain the original data bits. As the deinterleaving process only changes the order of received data, the error probability is intact. When passing through the CC-decoder and the RS-decoder, some errors may be corrected, which results in lower error rates [9].

### III. SIMULATION RESULTS

In this section, we have presented various BER vs. SNR plots for all the essential modulation and coding profiles in the standard on different channel models. We analyzed audio signal to transmit or receive data as considered for real data measurement. Figure 2, 3 and 4 display the performance on Additive White Gaussian Noise (AWGN), Rayleigh and Rician channel models respectively. The Bit Error Rate (BER) plot obtained in the performance analysis showed that model works well on Signal to Noise Ratio (SNR) less than 25 dB. Simulation results in figure 2 show the advantage of considering a ½ and 2/3 convolutinal coding and Reed-Solomon coding rate for each of the four considered digital modulation schemes (BPSK, QPSK, 4-QAM and 16-QAM). The performance of the system under BPSK modulation in 2/3 convolutional code rate is quite satisfactory as compared to other modulation techniques in AWGN channel.

found not to be suitable for transmission. It is also shown in this figure that the performance of QPSK and 4-QAM is found more better than BPSK modulation for a ½ convolutional code rate with respect to SNR values.

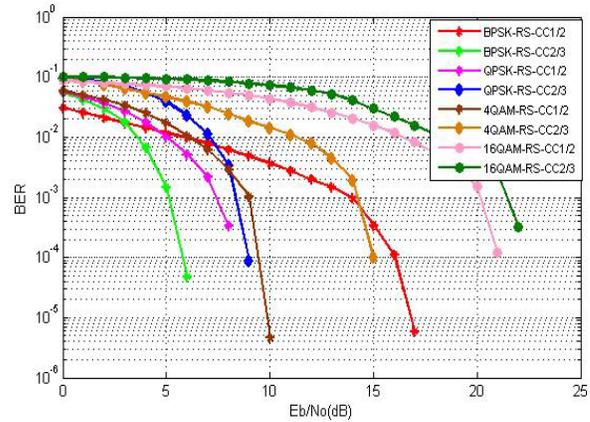

**Figure 3: System performance under different modulation schemes for a Convolutional Encoder with a 1/2 and 2/3 code rates in Rayleigh channel.**

In figure 4, it also shows that the BER performance of convolutional 2/3 code rate for BPSK modulation technique is better than all other modulation techniques and there is a little difference exists between BPSK-1/2 and BPSK-2/3 convolutional code rated.

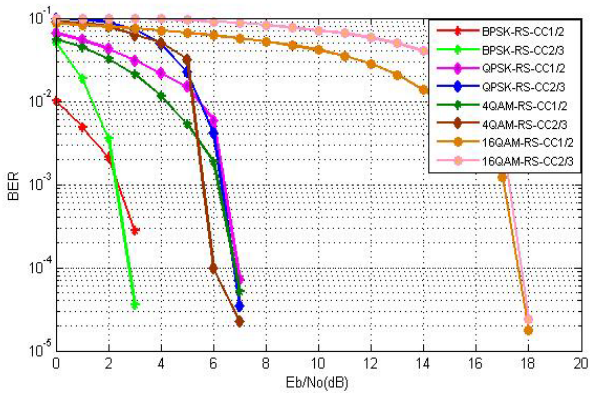

**Figure 2: System performance under different modulation schemes for a Convolutional Encoder with a 1/2 and 2/3 code rates in AWGN channel.**

The Bit Error Rate under BPSK modulation technique for a typical SNR value of 3 dB is .000035303 which is smaller than that of other modulation techniques.

In figure 3 with Rayleigh channel, the BER performance in case of 16-QAM modulation in 2/3 convolutional code rate is

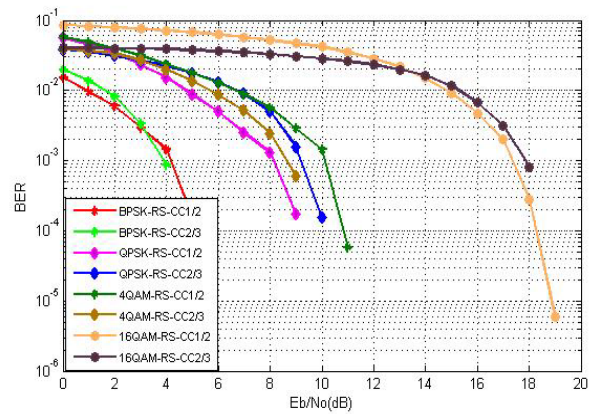

**Figure 4: System performance under different modulation schemes for a Convolutional Encoder with a 1/2 and 2/3 code rates in Rician channel.**

The transmitted and received audio signal for such a case corresponding with time and amplitude coordinates is shown in fig5.

techniques and coding scheme. The effects of the FEC (Forward Error Correction) and different communication channels were also evaluated in the form of BER. Performance results highlight the impact of modulation scheme and show that the implementation of an interleaved Reed-Solomon with 2/3 rated convolutional code under BPSK modulation technique under different channels provides satisfactory performance among the four considered modulations.

The IEEE 802.16 standard comes with many optional PHY layer features, which can be implemented to further improve the performance. The optional Block Turbo Coding (BTC) can be implemented to enhance the performance of FEC. Space Time Block Code (STBC) can be employed to provide transmit diversity.

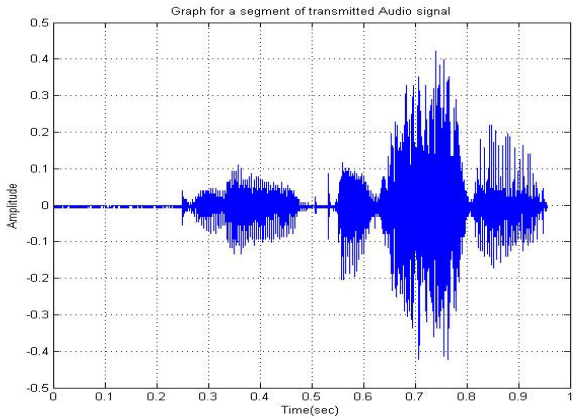

*(a)*

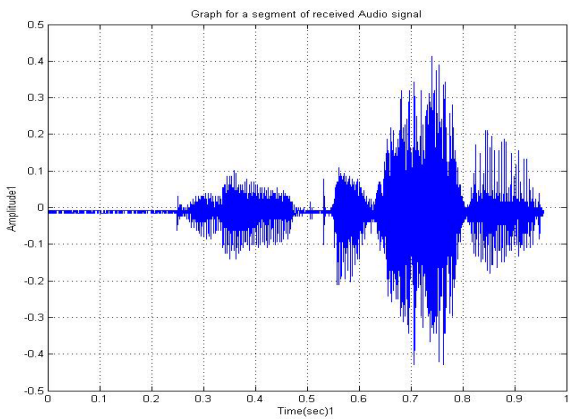

*(b)*

**Figure 5: A segment of an audio signal, (a) Transmitted (b) Retrieved**

## IV. CONCLUSION AND FUTURE WORKS

A performance analysis of an Wimax (Worldwide Interoperability for Microwave Access) system adopting concatenated Reed-Solomon and Convolutional encoding with block interleaver has been carried out. The BER curves were used to compare the performance of different modulation